\documentclass[journal]{IEEEtran}
\usepackage{xcolor}
\usepackage{cite}
\usepackage{amsmath,amssymb,amsfonts}
\usepackage{graphicx}
\usepackage{enumitem}
\usepackage{multirow}
\definecolor{Gray}{gray}{0.9} 
\definecolor{LightBlue}{RGB}{173, 216, 230} 
\usepackage{textcomp}
\usepackage{graphicx} 
\usepackage{booktabs} 
\usepackage{amsmath}
\usepackage{amssymb}
\usepackage{algorithm}
\usepackage{subcaption}
\usepackage{comment}
\usepackage{soul}          
\usepackage{makecell}
\usepackage{adjustbox}

\usepackage[noend]{algpseudocode} 
\algrenewcommand\algorithmiccomment[1]{\quad\{#1\}}

\usepackage{array} 
\usepackage{caption} 
\usepackage[justification=centering]{caption}
\def\BibTeX{{\rm B\kern-.05em{\sc i\kern-.025em b}\kern-.08em
    T\kern-.1667em\lower.7ex\hbox{E}\kern-.125emX}}

\makeatletter
\renewcommand{\alglinenumber}[1]{\footnotesize\arabic{ALG@line}:}
\makeatother

\begin{document}

\title{ReLATE+: Unified Framework for Adversarial Attack Detection, Classification, and Resilient Model Selection in Time-Series Classification}



\author{Cagla Ipek Kocal,
        Onat Gungor,
        Tajana~Rosing,
        and Baris Aksanli
\thanks{Cagla Ipek Kocal and Baris Aksanli are with the Department of Electrical and Computer Engineering, San Diego State University, San Diego (email: \{ckocal0169, baksanli\}@sdsu.edu).

Onat Gungor and Tajana Rosing are with the Department of Computer Science and Engineering, University of California San Diego, San Diego (email: \{ogungor, tajana\}@ucsd.edu). }
} 

\maketitle
\begin{abstract}
Minimizing computational overhead in time-series classification, particularly in deep learning models, presents a significant challenge due to the high complexity of model architectures and the large volume of sequential data that must be processed in real time. This challenge is further compounded by adversarial attacks, emphasizing the need for resilient methods that ensure robust performance and efficient model selection. To address this challenge, we propose ReLATE+, a comprehensive framework that detects and classifies adversarial attacks, adaptively selects deep learning models based on dataset-level similarity, and thus substantially reduces retraining costs relative to conventional methods that do not leverage prior knowledge, while maintaining strong performance. ReLATE+ first checks whether the incoming data is adversarial and, if so, classifies the attack type, using this insight to identify a similar dataset from a repository and enable the reuse of the best-performing associated model. This approach ensures
strong performance while reducing the need for retraining, and it generalizes well across different domains with varying data distributions and feature spaces. Experiments show that ReLATE+ reduces computational overhead by an average of 77.68\%, enhancing adversarial resilience and streamlining robust model selection, all without sacrificing performance, within 2.02\% of Oracle.
\end{abstract}




%
\section{Introduction}
\begin{figure}[t]
    \centering
    \includegraphics[width=0.48\textwidth]{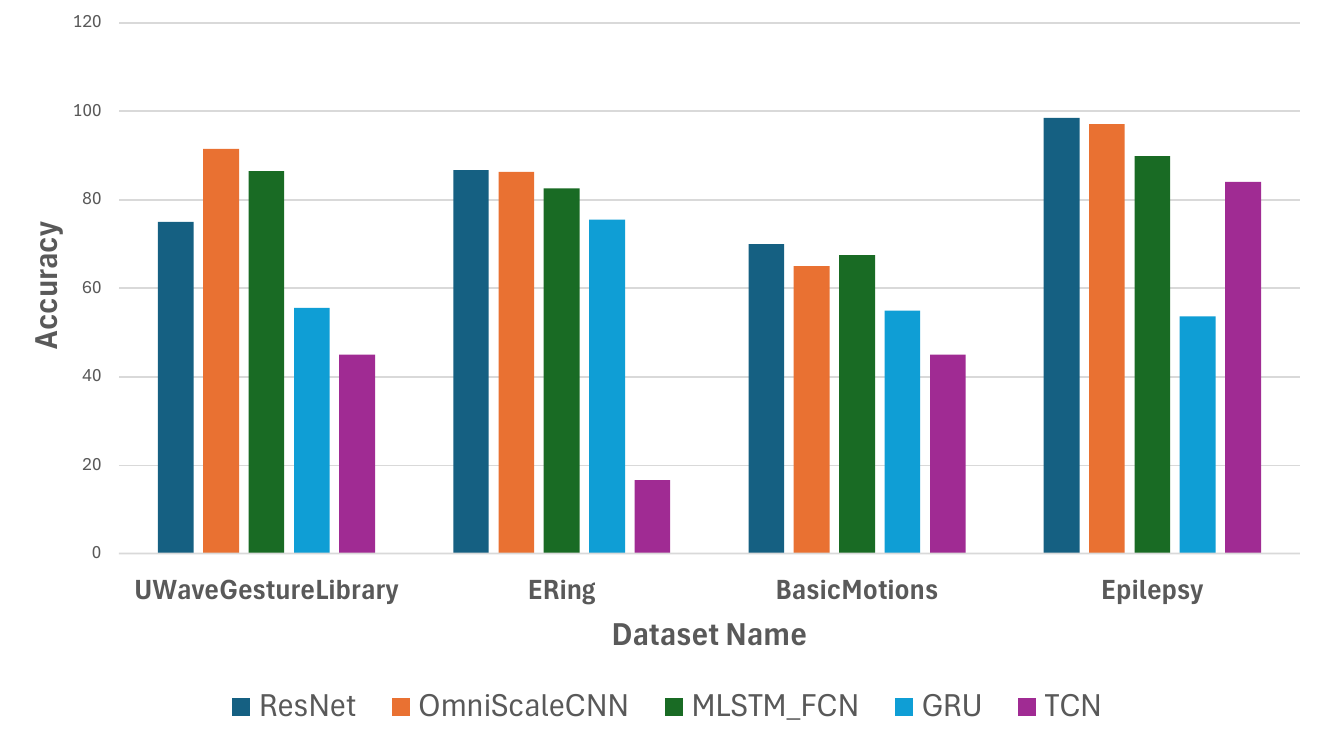}
    \caption{Performance of state-of-the-art deep learning models on multivariate time-series classification}
    \label{fig:model_performance}
\end{figure}

\IEEEPARstart{V}arious tasks rely on time-series data, i.e., sequences of observations collected over intervals,
including
anomaly detection \cite{gungor2024robust}, clustering \cite{holder2024review}, and classification \cite{ismail2019deep}. 
Among these tasks, time-series classification with machine learning (ML) has crucial use cases, e.g., network intrusion detection \cite{gungor2024rigorous}, event logs classification \cite{alzu2025cyberattack}, malware detection \cite{sayadi2021towards}, epileptic activity classification using EEG signals \cite{varli2023multiple}, smart agriculture using multispectral satellite imagery \cite{simon2022convolutional}, wearable activity recognition \cite{ordonez2016deep}, financial time-series classification \cite{liu2019multiscale}, and speech emotion recognition \cite{kim2017learning}, requiring robust, resilient, secure, and accurate ML-based solutions that can adapt in real time.
\\
\noindent
\indent Time-series ML applications face significant challenges due to the dynamic nature of streaming data, which is often limited or incomplete in real-time environments, making it impractical to wait for sufficient data accumulation to retrain models \cite{suarez2023survey}. Moreover, training ML models on new data is both computationally expensive and time-consuming, further complicating the process \cite{dempster2020rocket}. In this context, deep learning (DL) models are often favored for multivariate time-series classification tasks due to their ability to automatically extract relevant features. However, these DL models could show significant variability in classification performance, as shown in Figure~\ref{fig:model_performance}. These results highlight the substantial impact of model choice on classification outcomes, underscoring the critical need for careful and informed model selection in the context of DL-based time-series analysis. 
This motivates the need for efficient DL model selection methods that can adapt to new incoming data without requiring extensive retraining.

\noindent\indent 
Deep learning (DL) models are vulnerable to adversarial attacks, as they can be manipulated by small, imperceptible changes in the input data, especially when the data is limited or incomplete. These attacks introduce perturbations that obscure essential patterns, potentially leading to misclassifications in high-stakes applications where accuracy is critical.
For example, small perturbations in medical sensor data could lead to incorrect diagnoses, potentially endangering lives, while attacks in security systems could result in unauthorized access or compromised safety \cite{gungor2023adversarial}. 
Mitigating these risks requires methods that not only adapt efficiently to new incoming data but also demonstrate resilience against adversarial attacks. 
\\
\noindent\indent 
Building on the need for resilience in streaming environments, the ability to effectively detect adversarial manipulations and perform robust, attack-aware classification forms a foundational component of secure time-series systems~\cite{kocal2025relate}. Rather than relying solely on costly model retraining, next-generation frameworks must be capable of flagging suspicious perturbations as data arrive, determining the nature of any tampering, and then seamlessly invoking classifiers that can both label and be resilient against adversarial inputs. Unifying lightweight attack detection with adaptive model selection maintains real-time performance without heavy computational demands. This allows it to stay accurate and reliable even against adaptive adversarial strategies in critical time-series applications.
\\
\noindent
\indent 
We propose ReLATE+ to tackle the computational and retraining challenges in DL-based time-series classification with streaming data, particularly in the presence of adversarial attacks. ReLATE+ begins by flagging adversarially attacked inputs, introducing minimal computational overhead compared to standard inference and avoiding the costly retraining cycles required by conventional model-centric defenses. It then infers the attack type and leverages this information to identify a similar dataset from a repository, enabling the reuse of a pre-trained classifier that has previously demonstrated robustness under comparable data and attack conditions. This three-step pipeline eliminates the need for exhaustive testing and retraining across all models and datasets. 
ReLATE+ achieves substantial computational savings, averaging 77.68\% reduction in overhead, with strong performance, within 2.02\% of Oracle, providing a fast and efficient solution for time-series classification under both adversarial and non-adversarial conditions.


\section{Background and Related Work}


\subsection{Adversarial Attacks}
Adversarial attacks pose a significant threat to the reliability of machine learning models by introducing subtle, often imperceptible perturbations to input data that lead to incorrect predictions \cite{papernot2016transferability}. These attacks are typically categorized into white-box and black-box settings. White-box attacks assume full access to the model’s architecture, parameters, and gradients, enabling attackers to craft highly effective perturbations by directly exploiting internal model information. In contrast, black-box attacks operate with no knowledge of the model’s internal workings and rely solely on observable outputs to guide the adversarial manipulation \cite{karim2022adversarial}.

Costa et al. \cite{costa2024deep} offer a taxonomy of adversarial attacks. Among the white-box methods, Fast Gradient Sign Method (FGSM) generates adversarial examples in a single step by applying perturbations in the direction of the gradient’s sign to maximize the model’s loss. Its iterative variant, the Basic Iterative Method (BIM), applies smaller perturbations across multiple steps, resulting in stronger adversarial examples. Momentum Iterative Method (MIM) further improves BIM by adding momentum in gradient updates to avoid poor local minima, enhancing attack stability and success. DeepFool operates by approximating the classifier’s decision boundaries as linear hyperplanes and iteratively adjusting inputs to find the minimal perturbation needed for misclassification. Auto Projected Gradient Descent (AutoPGD) dynamically adapts step sizes in optimization, making it a robust and efficient iterative white-box attack. ElasticNet, a white-box method, integrates both L1 and L2 regularization, enabling the generation of sparse and interpretable perturbations, valuable in domains requiring minimally invasive changes. In the black-box setting, Boundary Attack initiates from a misclassified input and gradually reduces the perturbation while maintaining misclassification. This is especially effective when model gradients are unavailable, suitable for real-world systems with restricted access. 

\subsection{Multivariate Time-Series Classification (MTSC)}
MTSC assigns predefined labels to time-series data by analyzing the temporal patterns across multiple variables simultaneously \cite{hsieh2021explainable}. Traditional ML methods often struggle with raw time-series data due to its temporal structure and high dimensionality, needing specialized approaches \cite{lin2024higher}. Numerous ML algorithms are designed to enhance the scalability and predictive capabilities of models for time-series classification. 
Zheng et al. developed a framework with multi-channel deep convolutional neural networks in combination with Multilayer Perceptrons \cite{zheng2016exploiting}. Grabocka et al. introduced a shapelet-based method with supervised selection and online clustering \cite{grabocka2016fast}. Ruiz et al. proposed a method combining DL, shapelets, bag-of-words approaches, and independent ensembles \cite{ruiz2020benchmarking}. Baldán et al. employed feature-based methods with traditional classifiers \cite{baldan2021multivariate}. Fawaz et al. reviewed various DL models for time-series classification, including CNN-based architectures such as ResNet and FCN, and evaluated them also on multivariate benchmarks \cite{fawaz2019deep}. Despite significant advancements in MTSC, the computational overhead remains a critical limitation. Most existing approaches rely on exhaustive training procedures that require significant computational resources.

\subsection{Similarity-Based Selection Approaches}
Similarity analysis has a crucial role in ML and time-series analysis, offering a foundation for dataset comparison, feature selection, and model evaluation. 
Marks examined three measures of similarity for comparing two sets of time-series vectors, including the Kullback-Leibler divergence, the State Similarity Measure, and the Generalized Hartley Metric \cite{marks2013validation}. Bounliphone et al. introduced a statistical test of relative similarity to address challenges in model selection for probabilistic generative frameworks \cite{bounliphone2015test}. Assegie et al. proposed a feature selection method using dataset similarity to improve the classification performance \cite{assegie2023multivariate}. Cazelles et al. introduced Wasserstein-Fourier distance 
to measure the dissimilarity between stationary time series \cite{cazelles2020wasserstein}. Xu et al. further explored Dynamic Time Warping as a robust method for time-series curve similarity, highly effective in handling variations in temporal alignment \cite{xie2020dtw}. Existing methods are tailored for static datasets and do not account for the dynamic nature of time-series data, lacking the adaptability required for evolving data and resilience against adversarial attacks.

\subsection{Adversarial Attacks in MTSC}
The streaming, high-dimensional, and time-sensitive nature of MTSC makes it especially susceptible to even minor adversarial perturbations, motivating a dedicated review of attacks in this domain. Several studies analyze the vulnerabilities of MTSC models to adversarial attacks and propose solutions to enhance resilience. 
Harford et al. adapt Adversarial Transformation Network on a distilled model, and show that 1-NN Dynamic Time Warping and Fully Convolutional Networks are highly vulnerable \cite{harford2020adversarial}. Galib et al. analyze time-series regression and classification performance under adversarial attacks, finding that Recurrent Neural Network models were highly susceptible \cite{galib2023susceptibility}.  
Siddiqui et al. proposed a regularization-based defense \cite{siddiqui2020benchmarking}. Gungor et al. developed a resilient stacking ensemble learning-based framework against various adversarial attacks \cite{gungor2022stewart}. However, these approaches overlook significant retraining or exhaustive experimentation whenever new attack scenarios or datasets emerge.


\subsection{Adversarial Attack Detection}
To address the retraining overhead of model-centric defenses, many works shifted toward attack detection, identifying manipulated inputs so that classifiers can apply specialized handling or fallbacks to improve robustness.\cite{metzen2017detecting}. Feature Squeezing reduces input complexity, such as bit-depth or resolution, and flags samples based on inconsistent predictions across transformations \cite{xu2018feature}. Mahalanobis distance based detection models the distribution of deep features per class and classifies inputs as adversarial if they fall outside high-density regions \cite{lee2018simple}. In time-series domains, RADAR applies a bi-directional LSTM autoencoder to detect perturbations through reconstruction error and uncertainty estimation \cite{harford2020radar}. TRAIT takes a spectral approach by analyzing the loss landscape during training to detect adversarial patterns with high generalizability \cite{xu2022pixels}. ReLATE+ builds on these approaches by integrating Fourier and Wavelet-based anomaly detection with similarity-driven model selection, enabling resilient performance without the need for costly retraining.

\noindent
\indent ReLATE+ addresses both adversarial robustness and retraining overhead by combining a targeted attack detection pipeline with similarity-driven model selection. It begins by screening incoming time-series data for adversarial perturbations, and when detected, classifies them into specific adversarial attack types. Using a pre-evaluated model repository, ReLATE+ computes dataset level similarity and dynamically selects the most effective classifier, eliminating the need for exhaustive DL model retraining. This not only adapts to the dynamic nature of new time-series data but also enhances robustness against adversarial attacks by prioritizing resilient model selection.


\begin{figure*}[t]
    \centering
    \includegraphics[trim=5 10 5 10, clip, width=0.9\textwidth]{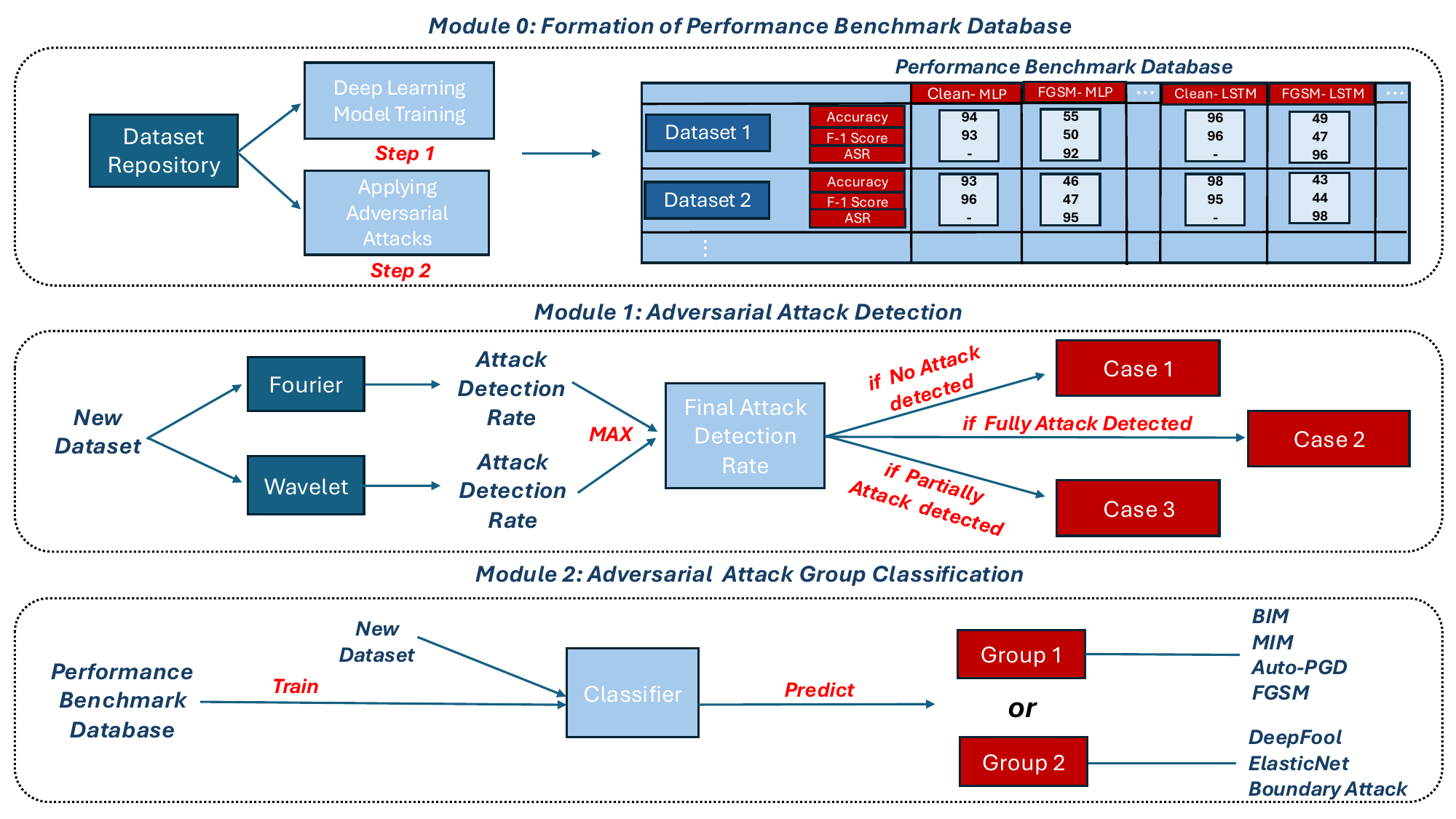}
    
    \vspace{1em} 

    \includegraphics[trim=5 3 5 3, clip, width=0.9\textwidth]{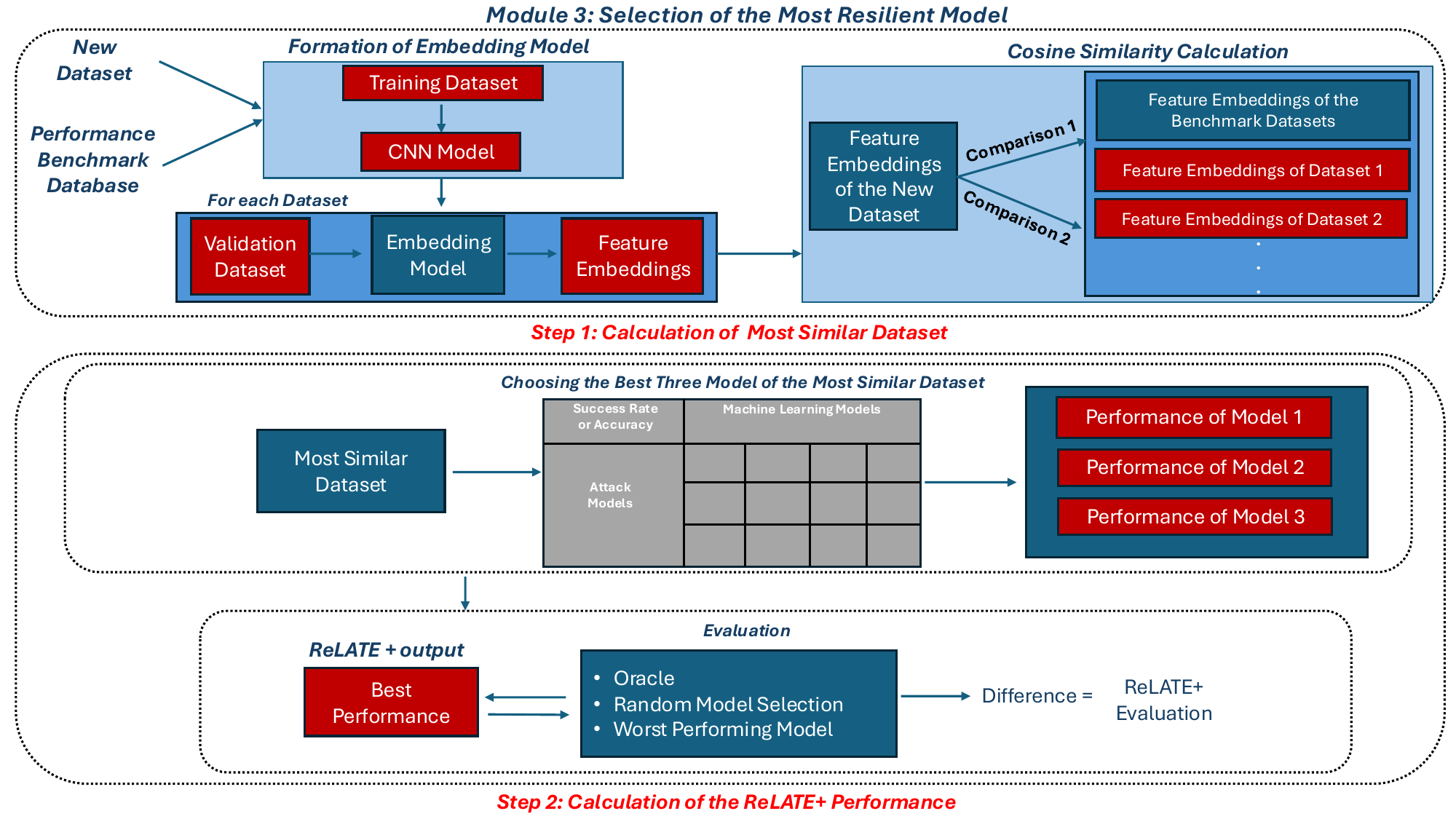}
    
    \caption{\protect ReLATE+ Framework}
    \label{fig:ReLATE+}
\end{figure*}


\section{Proposed Framework: ReLATE+}
ReLATE+ provides a robust approach for model selection in time-series classification, balancing resilience with computational efficiency. Its design is driven by the need to (1) avoid redundant model retraining through strategic reuse of existing models, (2) detect and adapt to adversarial threats in a lightweight manner, and (3) match incoming data to historically similar conditions to inform model selection. The architecture of ReLATE+, comprising four key modules, is illustrated in Figure \ref{fig:ReLATE+}. Module 0 constructs a Performance Benchmark Database (PBD) that stores pre-evaluated metrics for various DL models across multiple datasets and attack scenarios. Once new data is received, Module 1 classifies it as clean, partially attacked, or fully attacked. If an attack is detected, Module 2 further identifies the attack group. Module 3 matches the incoming data to the most similar dataset in the repository using a similarity metric, and selects the top performing models accordingly, eliminating exhaustive model evaluations. 

\subsection{Module 0: Formation of Performance Benchmark Database}
This module constructs a PBD using previously analyzed datasets. It is conducted offline prior to receiving any incoming data and consists of two main steps: deep learning model training and adversarial attack generation.

\subsubsection{Step 1: Deep Learning Model Training}
This step trains our DL models on various datasets. We begin with hyperparameter tuning to identify the optimal settings for each model and dataset. We then train the models and record their performance (accuracy and F1-score) in the PBD. We use the training dataset for model training, the validation dataset for hyperparameter tuning, and reserve the test dataset to evaluate the final performance. This module processes the input datasets and generates the clean data (no attack) performance. The DL models span a diverse range, each addressing specific challenges of time-series data, including capturing long-range temporal dependencies, identifying cyclical patterns within sequential data, and managing high dimensionality in multivariate settings. We select 14 state-of-the-art DL models: MLP \cite{wang2017time}, FCN \cite{wang2017time}, ResNet \cite{wang2017time}, LSTM \cite{karim2017lstm}, GRU \cite{karim2017lstm}, LSTM-FCN \cite{karim2017lstm}, GRU-FCN \cite{karim2017lstm}, MSWDN \cite{elsayed2018deep}, TCN \cite{bai2018empirical}, MLSTM-FCN \cite{karim2019multivariate}, InceptionTime \cite{ismail2020inceptiontime}, Residual CNN \cite{zou2019integration}, OmniScaleCNN \cite{tang2020omni}, and Explainable Convolutional Model \cite{fauvel2021xcm}.\\               
\subsubsection{Step 2: Applying Adversarial Attacks}
In this step, each dataset-model pairing undergoes evaluation through nine state-of-the-art adversarial attacks, both white-box and black-box. White-box attacks have full access to the model’s architecture and parameters to generate adversarial examples, while black-box attacks rely solely on the model's outputs, without knowledge of its internal structure. 
This module generates adversarial attack versions of each dataset, testing the robustness of DL models against various attack types. The results (accuracy, F1-score, and attack success rate) are recorded in the PBD. Adversarial attacks are applied to the test portion of the data using models trained in Step 1 with optimized hyper-parameters. Each attack is chosen for its unique approach to disrupting data and exposing model weaknesses, from simple gradient-based methods to complex iterative strategies. We select seven white-box and black-box attacks \cite{costa2024deep}: Fast Gradient Sign Method (FGSM), DeepFool, Basic Iterative Method (BIM), Momentum Iterative Method (MIM), ElasticNet, Auto Projected Gradient Descent (AutoPGD) and Boundary Attack.  

\subsection{Module 1: Adversarial Attack Detection}
In real-world scenarios, adversarial inputs often impact only portions of the data, rendering simple binary classification (attack vs. no attack) inadequate for capturing varying degrees of compromise. To better reflect this complexity, we classify incoming data as clean (Case 1), fully attacked (Case 2), or partially attacked (Case 3). For this classification task, we apply both Wavelet Transform and Fourier Transform, selected for their complementary strengths in detecting different types of adversarial perturbations. The structure of incoming time-series data, whether perturbations are localized or spread throughout, is typically unknown. The Wavelet Transform excels at identifying short-term, high-frequency anomalies common in gradient-based attacks by analyzing both time and frequency domains \cite{zhang2024frequency}. In contrast, the Fourier Transform is suited for capturing global frequency shifts or consistent periodic distortions introduced by systematic or optimization-based attacks \cite{kiruluta2025hybrid}. Together, these transforms enable robust detection across a broad range of perturbation patterns. To support classification into the three cases, we compute the Attack Detection Rate, quantifying the proportion of adversarial perturbation present in the input.


After obtaining detection scores from both the Wavelet and Fourier Transforms, the higher of the two is selected to represent the final attack detection ratio. This approach ensures that even localized or subtle adversarial signals, possibly detected by only one method, are not overlooked. Given the unpredictable nature of adversarial perturbations, favoring the stronger signal helps safeguard against misclassifying compromised data as clean, which could otherwise lead to inappropriate model selection or reduced robustness. In cases where the two methods disagree significantly, selecting the higher value prioritizes caution, acknowledging that an undetected attack can be more harmful than responding to a false alarm.

The selected attack detection ratio is then used to classify the data into one of three categories. This three-way classification is necessary to capture the full spectrum of real-world scenarios: some data may be entirely clean (Case 1), others consistently attacked (Case 2), while many fall in between where only a portion is adversarially perturbed (Case 3). Let $\mathbf{T}$ be a predefined threshold: if the ratio is below $\mathbf{T}$, the data is labeled clean; if it exceeds $\mathbf{1-T}$, it is labeled fully attacked; otherwise, it is considered partially attacked.


\subsection{Module 2: Adversarial Attack Group Classification}
When adversarial attack is detected in Module 1, the next step is to determine the nature of the attack. Accurately identifying the type of perturbation enables more targeted and effective model selection in later stages, as different attacks can degrade model performance in distinct ways. Rather than pinpointing the exact attack, this module categorizes the input into one of two broader adversarial attack groups based on shared structural characteristics. Group 1: Iteration-Based Attacks consists of gradient-based attacks (including FGSM, BIM, MIM, and AutoPGD) which typically generate small, directionally consistent perturbations by directly leveraging model gradients. Group 2: Optimization- and Decision-Based Attacks comprises optimization-based and decision-based attacks such as DeepFool, Boundary Attack, and ElasticNet, which rely on iterative optimization, decision boundary exploration, or sparse modifications. A classifier trained on the PBD is used to distinguish between these groups, enabling the system to make context-aware decisions based on the likely nature of the attack.

\subsection{Module 3: Most Resilient Model Selection}
\subsubsection{Step 1: Most Similar Dataset Calculation}
When a new dataset arrives, mimicking real-time conditions, ReLATE+ initiates similarity comparison for model selection. The process begins by training a lightweight DNN-based similarity function, built on a simple CNN architecture, using the training portion of each dataset in the PBD. Once trained, we use these CNNs to extract feature embeddings from the validation portions of their respective datasets. For the incoming dataset, we train a separate similarity function using the training portion of its data and extract embeddings from the validation portion. We then normalize these embeddings to ensure efficient quantification of average similarity between datasets based on their feature distributions. We compare the embeddings extracted from the incoming dataset to those in the PBD with an Embedding Similarity Metric, e.g., Cosine similarity.\\  

\subsubsection{Step 2:  ReLATE+ Performance Assessment}
Here, we find the dataset with the highest similarity score and select its top three performing DL models (ranked by test performance recorded in Module 0: Step 1 and 2). This approach minimizes computational overhead by eliminating the need to retrain or test all models on the incoming dataset. Next, we assess the performance of the selected models on the new dataset by training them on its training portion and evaluating their accuracy and resilience on the test portion.
For clean data, we measure performance with accuracy. For the adversarial data, we assess resilience with attack success rate (ASR). This selection is directly informed by the similarity analysis, ensuring that the chosen model is well-suited to handle the unique characteristics and potential adversarial challenges of the new data. 
The best model is then deployed on the new dataset for real-time use. We compare the performance of the best model with oracle, random model selection, and the worst-performing model to evaluate ReLATE+ output.

\begin{figure}[t]
  \centering
  \begin{subfigure}[b]{\columnwidth}
    \includegraphics[trim=70 200 70 20, clip, width=\textwidth]{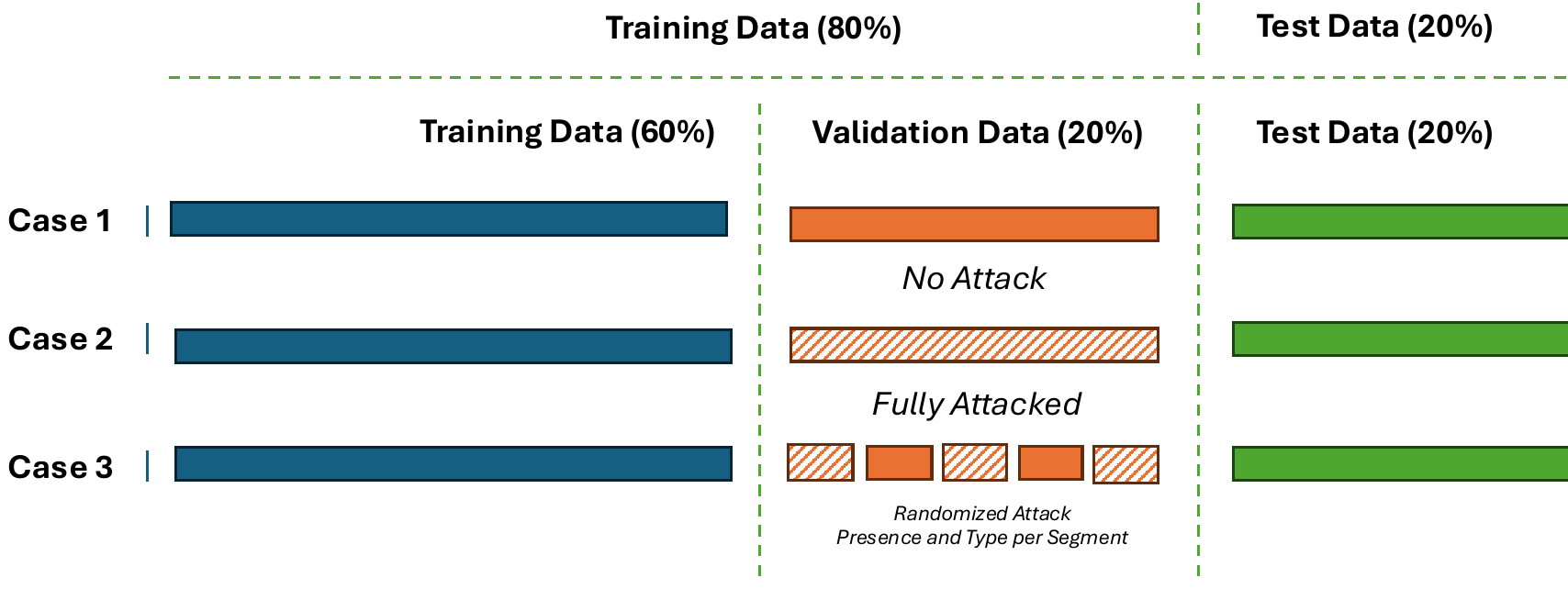}
    \caption{Incoming Data Setup}
    \label{fig:incoming-setup}
  \end{subfigure}

  \vspace{1em} 

  \begin{subfigure}[b]{\columnwidth}
    \includegraphics[trim=5 50 10 3, clip, width=\textwidth]{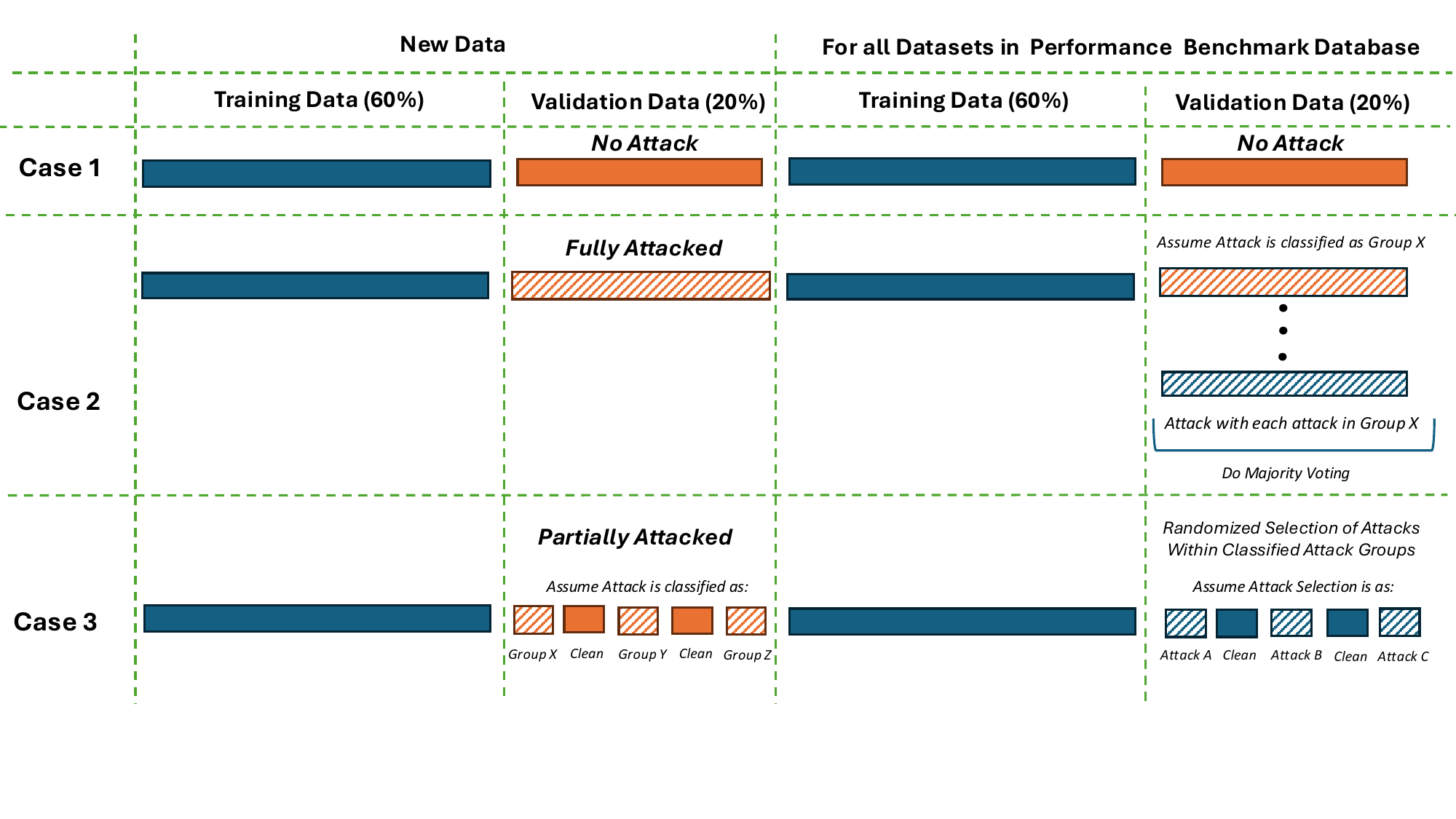}
    \caption{Case Evaluation Explanation}
    \label{fig:case-evaluation}
  \end{subfigure}

  \caption{Overall Case Explanations}
  \label{fig:cases}
\end{figure}
\subsection{Using ReLATE+ on a New Dataset and Design Choices}
\subsubsection{Incoming Data Setup} We conduct a series of experiments where each of the datasets in our PBD is treated as a new, unseen arrival in rotation. In each round, one dataset is designated as the ``new arrival'', while the remaining datasets are treated as the pre-existing ``drive'' datasets. For each drive dataset, we assess key performance metrics, including accuracy, F1-score, and ASR, across all models, encompassing both the clean and adversarially attacked variants. 
Figure \ref{fig:cases} part a illustrates the training-validation splits for each case. In all cases, the training dataset is split into an 80\% training set and a 20\% validation set to facilitate model training and validation for comparisons. This split is determined after several trials to find the most effective training-validation ratio. The custom CNN is trained using the training portion of the dataset, while the validation portion is utilized for similarity measurement during case implementation. Overall, we design three cases to assess model selection under different adversarial conditions:

\noindent \textbf{Case 1: No Adversarial Attack –} When the incoming dataset is classified as clean (Figure \ref{fig:cases}a), it is compared with the clean versions of each dataset in the drive (Figure \ref{fig:cases}b). We find the most similar dataset in the drive and select its three best-performing models. We evaluate the selected models on the incoming dataset to identify the best among them. 

\noindent \textbf{Case 2: Full Attacks – } When the incoming dataset is identified as being subject to a single type of adversarial attack (Figure \ref{fig:cases}a), a dataset similarity analysis is performed using a majority voting approach to find the most similar dataset. If the dataset is classified as fully attacked by Module 1 (Case 2), its corresponding attack group is identified by Module 2. With this attack group information, each dataset in the PBD is attacked using all individual attacks within the predicted group. For each attack, a similarity analysis is conducted between the resulting dataset and the incoming dataset, and the results are recorded. After all comparisons are completed, majority voting is applied to determine the most similar dataset overall (Figure \ref{fig:cases}b). This strategy addresses the uncertainty regarding the exact attack affecting the incoming dataset by evaluating all plausible candidates within the predicted group. Once the most similar attacked dataset is identified, its top three models—ranked by attack success rate (ASR), where a lower ASR indicates higher model resilience—are selected and evaluated on the incoming dataset.

\noindent \textbf{Case 3: Partial Attacks –} We design a randomized attack scenario to simulate diverse and unpredictable adversarial conditions. We divide each dataset's validation portion into five equal segments. For each segment, a binary decision is made at random to determine whether the segment will be subjected to an adversarial attack. If an attack is to be applied, the attack type is randomly selected from the predefined set of adversarial methods described previously (Figure \ref{fig:cases}a).

Once the attack pattern across the five segments is established and treated as the incoming dataset, Module 1 classifies it as partially attacked (Case 3). Then, each segment is isolated and re-evaluated individually by Module 1 to distinguish between clean and attacked portions. Segments identified as attacked are then passed to Module 2, where their attack group is classified according to the framework. The sequence of clean and attacked segments—along with the corresponding attack groups for the attacked segments—is recorded to mirror the structure of the incoming dataset across the datasets in the PBD. The validation portion of each benchmark dataset is similarly divided into five segments. Segments identified as clean remain unchanged, while those classified as attacked are perturbed using a randomly selected attack from the corresponding predicted attack group. This approach preserves the structural integrity of the datasets while introducing controlled attack variation aligned with the incoming pattern (Figure \ref{fig:cases}b). After generating these patterned datasets, the most similar dataset is identified based on the established segment-level structure, and the subsequent steps follow the procedure outlined in the previous cases.

  



\begin{algorithm}[t]
  \caption{Similarity Driven Dataset Selection}
  \label{alg:similarity}

  \begin{algorithmic}[1]       
  \Require Input $D_{\text{in}}$, Performance Benchmark Database $\mathcal{B}$
  \Require Modules
          $\textsc{M1}$ (detector),
          $\textsc{M2}$ (attack–group classifier),
          $\textsc{M3-1}$ (similarity),
          $\textsc{M3-2}$ (performance)

  \State $c \gets \textsc{M1}(D_{\text{in}})$
  \State $\textit{val}_{\text{in}} \gets \text{ValidationSplit}(D_{\text{in}})$
  \State $\textit{test}_{\text{in}} \gets \text{TestSplit}(D_{\text{in}})$
  \If{$c = \textsc{Case1}$}      \Comment{Dataset fully clean}
      \ForAll{$\mathcal{D}\in\mathcal{B}$}
          \State $\textit{val}_{\mathcal{D}} \gets \text{ValidationSplit}(\mathcal{D})$
      \EndFor
      \State $\textit{sim}  \gets \textsc{M3-1}(\textit{val}_{\text{in}},\textit{val}_{\mathcal{D}})$
      \State $\textit{perf} \gets \textsc{M3-2}(\textit{sim},\textit{test}_{\text{in}})$
      \State \Return $(\textit{sim},\textit{perf})$
  \EndIf
  \If{$c = \textsc{Case2}$} \Comment{single unknown attack}
      \State $\mathcal{G} \gets \textsc{M2}(D_{\text{in}})$
      \ForAll{$a\in\mathcal{G}$}
          \ForAll{$\mathcal{D}\in\mathcal{B}$}
              \State $\textit{val}_{\mathcal{D}}
                     \gets \text{ApplyAttack}\bigl(a,\text{ValidationSplit}(\mathcal{D})\bigr)$
          \EndFor
          \State $\textit{sim}[a] \gets \textsc{M3-1}(\textit{val}_{\text{in}},\textit{val}_{\mathcal{D}})$
      \EndFor
      \State $\textit{sim}_{\text{best}} \gets \text{MajorityVoting}(\textit{sim})$
      \State $\textit{perf} \gets \textsc{M3-2}(\textit{sim}_{\text{best}},\textit{test}_{\text{in}})$
      \State \Return $(\textit{sim},\textit{perf})$
  \EndIf
  \If{$c = \textsc{Case3}$} \Comment{mixed clean / attacked}
      \State $k \gets 5$ \Comment{number of segments}
      \For{$i \gets 1$ \textbf{to} $k$}  \Comment{detect attacked segments}
          \State $s_i \gets$ $i$-th segment of $\textit{val}_{\text{in}}$
          \State $\textit{status}[i] \gets \textsc{M1}(s_i)$
          \If{$\textit{status}[i] = \textsc{Attacked}$}
              \State $\mathcal{G}[i] \gets \textsc{M2}(s_i)$
          \Else
              \State $\mathcal{G}[i] \gets \varnothing$
          \EndIf
      \EndFor

      \ForAll{$\mathcal{D}\in\mathcal{B}$} \Comment{create attacked replicas}
          \State $\textit{val}_{\mathcal{D}} \gets \text{ValidationSplit}(\mathcal{D})$
          \Statex\hspace{\algorithmicindent}\textbf{Partition} $\textit{val}_{\mathcal{D}}$ into $t_1,\dots,t_k$
          \For{$i \gets 1$ \textbf{to} $k$}
              \If{$\textit{status}[i] = \textsc{Attacked}$}
                  \State $a \gets \text{RandomSelect}(\mathcal{G}[i])$
                  \State $t_i \gets \text{ApplyAttack}(a,t_i)$
              \EndIf
          \EndFor
          \State $\textit{val}_{\mathcal{D}} \gets \bigcup_{i=1}^{k} t_i$
      \EndFor

      \State $\textit{sim}  \gets \textsc{M3-1}(\textit{val}_{\text{in}},\textit{val}_{\mathcal{D}})$
      \State $\textit{perf} \gets \textsc{M3-2}(\textit{sim},\textit{test}_{\text{in}})$
      \State \Return $(\textit{sim},\textit{perf})$
  \EndIf
  \end{algorithmic}
\end{algorithm}

The complete workflow in this section is concisely captured in Algorithm~\ref{alg:similarity}, where each step is expressed in high-level pseudo-code to highlight the sequence of operations.

\subsubsection{Design Choices}
All experiments in this paper were conducted with $\epsilon = 0.1$ for attack methods that allow adjustable perturbation strength, as this value offers a balanced trade-off between attack effectiveness and subtlety.

To implement the adversarial attack detection strategy in Module 1, the following design choices are introduced to implement the detection strategy. An attack detection method is trained on transformed clean training data by modeling normalized deviation scores, and it flags anomalies when these scores exceed a threshold based on a high-percentile value. To determine the threshold for classifying input data as Clean (Case 1), Partially Attacked (Case 3), Fully Attacked (Case 2), a series of experiments is conducted on the validation sets within the PBD. The Fourier and Wavelet-based detection methods are trained using the corresponding training partitions of each dataset. The threshold is then chosen to achieve an optimal balance between false positive and false negative rates, as observed across the validation evaluations. Based on this analysis, a threshold of \( T = 13\% \) is selected to maximize overall attack detection performance. Specifically, if the detection rate is below 13\%, the data is considered clean (Case 1); if it falls between 13\% and 87\%, it is classified as partially attacked (Case 3); and if it exceeds 87\%, the dataset is labeled as fully attacked (Case 2). For Case 3, since the incoming data arrives in five equally sized chunks, the possible degrees of partial attack correspond to chunk-level perturbations, i.e., 20\%, 40\%, 60\%, and 80\%. If the overall detection rate falls within the 13\%–87\% range, indicating a partial attack, the system selects the closest predefined intensity level (within a ±10\% tolerance) to estimate the proportion of attacked chunks. After this, as each segment is independently processed through Module 1 to determine whether it is attacked or clean, a threshold of 50\% is applied to each, reflecting the two possible outcomes.

The following design choices are introduced for Module 2: Adversarial Attack Group Classification, which is responsible for categorizing detected attacks into predefined groups based on their characteristics. For adversarial attack classification, XGBoost\cite{chen2016xgboost} is selected as the classifier due to its strong predictive performance compared to other supervised learning methods. The model is trained using datasets from the PBD, where each dataset is subjected to all available adversarial attack types. Samples are then labeled as either Group 1: Iteration-Based Attacks or Group 2: Optimization- and Decision-Based Attacks, depending on which attack group the applied attack belongs to, forming the labeled training data for the classification task. When an incoming dataset is classified as fully attacked (Case 2), the trained classifier model is used to directly predict the corresponding attack group. For partially attacked datasets (Case 3), the specific attacked segments identified in Module 1 are isolated, and the classifier is then applied to those segments to determine the most likely attack group.

\section{Experimental Setup}
\noindent \textbf{Dataset Description:}
We utilize six datasets from the UEA Multivariate Time-Series Classification (MTSC) archive \cite{uea2018}. The selected datasets span two domains: five are related to Human Activity Recognition (HAR) and one is derived from motion data recordings. This cross-domain selection is intended to evaluate the robustness and generalizability of the proposed method across diverse application areas. Multivariate datasets are prioritized since they reflect the complexity of real-world time-series problems, where multiple interrelated variables must be considered together to accurately capture underlying temporal patterns \cite{ruiz2021bakeoff}. As summarized in Table~\ref{table:uea_datasets}, the datasets vary in training size, test size, number of dimensions, sequence length, and class count, allowing for comprehensive evaluation under varying data characteristics.


\begin{table}[]
\centering
\caption{Selected datasets from the UEA repository}
\label{table:uea_datasets}
\resizebox{\columnwidth}{!}{ 
\begin{tabular}{cccccc}
\hline
\textbf{Dataset}  & \textbf{Train} & \textbf{Test} & \textbf{Dim.} & \textbf{Len.} & \textbf{Classes} \\ \hline
NATOPS            & 180            & 180           & 24            & 51            & 6                \\
UWaveGestureLibrary   & 120           & 320          & 3             & 315           & 8                \\
Cricket           & 108            & 72            & 6             & 1197          & 12               \\
ERing             & 30             & 270           & 4             & 65            & 6                \\
BasicMotions      & 40             & 40            & 6             & 100           & 4                \\
ArticularyWordRecognition      & 275            & 300           & 9             & 144            & 25                \\
\hline
\end{tabular}
}
\end{table}


\noindent \textbf{Hardware Setup:}
We use a PC equipped with an Intel Core i7-9700K CPU (8 cores), 32 GB of RAM, and a 16 GB NVIDIA GeForce RTX 2080 dedicated GPU. 

\noindent \textbf{Evaluation Metrics:}
We use three metrics: accuracy, F1-score, and attack success rate (ASR). Accuracy measures the proportion of correctly classified instances out of the total number of samples. F1-score evaluates the balance between precision and recall, making it well-suited for datasets with class imbalance. ASR measures the effectiveness of adversarial attacks by calculating the percentage of instances where model predictions are successfully altered.



\noindent \textbf{Dataset Similarity Calculation:}
We quantify dataset similarity using a custom CNN with two 1D convolutional layers, adaptive max-pooling, and dropout to extract features from both clean and attacked data. The final fully connected layer maps these features to class predictions. The resulting embeddings are then normalized using L2 normalization to ensure consistency and scale invariance. We use cosine similarity between the embeddings, which measures the angular similarity between two vectors in a multi-dimensional space:
\[
\text{Cosine Similarity} = \frac{\mathbf{A} \cdot \mathbf{B}}{\|\mathbf{A}\| \|\mathbf{B}\|}
\]

where \(\mathbf{A}\) and \(\mathbf{B}\) are the embedding vectors. Cosine similarity ranges from \(0\) (orthogonal) to \(1\) (identical). To identify the most appropriate similarity metric for ReLATE+, we also evaluate performance using DTW \cite{Petitjean2011}, which aligns sequences by minimizing temporal distortions, and Wasserstein Distance \cite{Flamary2017}, which quantifies distributional differences by computing the optimal transport cost between probability distributions. Based on the superior performance of the custom CNN combined with cosine similarity, we select it as the primary similarity metric (see Section V.C and Fig. 5).

\noindent \textbf{Baselines:}
We compare ReLATE+ against three baseline approaches: random model selection, Oracle (best model) and the worst-performing model. All baseline comparisons are conducted using the same set of 14 deep learning models (see Module 0: Formation of PBD Step 1), ensuring consistency and fairness across evaluations.
Random model selection uses a Monte Carlo approach, where an ML model is randomly chosen for each dataset. This process is repeated 1,000 times, with the average performance calculated as the baseline score for random selection. Oracle represents the maximum accuracy recorded on the test data for each dataset among all models. Oracle reflects the upper performance bound achievable by the optimal model with exhaustive search and is usually computationally impractical.
The worst-performing model represents the least effective model among all the possible DL models in the database, reflecting the lower performance bound.


For the state-of-the-art comparison, we conceptualize it as a spectrum bounded by Oracle and Random Model Selection. The appropriate choice within this spectrum depends on the specific priorities and the available model space. If the application demands strict high performance—such as maximizing accuracy or ensuring strong adversarial resilience—regardless of computational cost, Oracle represents the ideal, though computationally expensive, option. In contrast, if computational efficiency is the primary concern and performance is of lesser importance, Random Model Selection offers a lightweight alternative. However, for scenarios where both strong performance and computational efficiency are critical, ReLATE+ provides the most balanced and practical solution.

\section{Results\\}

\subsection{ReLATE+ Performance}
This section presents the outcomes of Module 1: Adversarial Attack Detection, Module 2: Adversarial Attack Group Classification, and Module 3: Most Resilient Model Selection, followed by a detailed overhead analysis, and an ablation study. For Case 1 and Case 2, all possible scenarios were evaluated. However, as Case 3 involves a stochastic process in which each data segment is randomly assigned to be either clean or attacked—and if attacked, an attack type is also randomly selected—five representative random scenarios were generated for evaluation due to the inherent randomness and combinatorial explosion of possible configurations.

\subsubsection{Module 1: Adversarial Attack Detection Results}
The objective of Module 1 is to classify incoming data into one of three categories: Case 1 (no attack), Case 2 (fully attacked), or Case 3 (partially attacked). The results from Module 1 demonstrate that our technique effectively distinguishes between clean, partially attacked, and fully attacked datasets.

\begin{table}[ht]
\centering
\scriptsize
\begin{adjustbox}{width=\columnwidth}
\begin{tabular}{lrr}
\toprule
\textbf{Dataset} & \textbf{Fourier (\%)} & \textbf{Wavelet (\%)} \\
\midrule
UWave                     & 7.50   & 5.60   \\
ERing                     & 12.22  & 12.20  \\
BasicMotions              & 100.00 & 0.00   \\
Cricket                   & 6.90   & 4.17   \\
NATOPS                    & 12.22  & 9.44   \\
ArticularyWordRecognition & 3.66   & 3.33   \\
\bottomrule
\end{tabular}
\end{adjustbox}
\caption{Attack detection results for Case 1}
\label{tab:module_1_case_1}
\end{table}

Table~\ref{tab:module_1_case_1} presents the results for Case 1: No Adversarial Attack. In the clean data setting, a lower attack detection rate is desirable, as it indicates the absence of false positives. These findings show that all but one dataset in the PBD were correctly classified as clean, with Fourier and Wavelet complementing each other, Fourier capturing global patterns, and Wavelet detecting localized variations, to ensure robustness. An exception occurred with the BasicMotions dataset. While the Wavelet Transform correctly identified the data as clean, the Fourier Transform produced a high attack detection rate. Since the final decision is based on the maximum detection score, this led to a misclassification. 

For Case 2, 40 out of 42 dataset-attack combinations (derived from 6 datasets and 7 attacks) were correctly identified as fully attacked using the combined approach, as shown in Table~\ref{tab:fw_case2_comparison}. In the table, the first value in each cell corresponds to the detection rate from the Fourier Transform, the second from the Wavelet Transform, and the bolded value represents the maximum of the two, which is used for final classification. The results reflect strong overall performance, with only two misclassifications across all evaluations. For the NATOPS dataset under FGSM, BasicIterative, MIM, and AutoPGD attacks, Wavelet achieved moderate detection rates (50–60\%), which resulted in misclassification as Case 3, since this range falls within the 13–87\% detection threshold defined for partial attacks. In contrast, Fourier consistently produced higher detection rates (95–100\%), correcting Wavelet’s output and underscoring the benefit of their combined use. Also in the case of DeepFool attacks on UWave and ERing, Wavelet demonstrated limited performance (41.25\% and 57.03\%, respectively), but Fourier compensated with significantly higher detection rates (87.71\% and 100\%). Similarly, Fourier failed to detect any attacks on the BasicMotions dataset, whereas Wavelet successfully corrected the classification. Since final decision is based on the maximum score between the two methods, these results further highlight their complementary strengths. This complementary behavior arises from the distinct characteristics of the two transforms: Fourier is more sensitive to global frequency-domain perturbations, making it effective for attacks that induce consistent spectral distortions, while Wavelet excels at capturing localized, transient anomalies in both time and frequency domains. However, for DeepFool attacks on NATOPS and Cricket, both Fourier and Wavelet achieved only around 50\% detection accuracy, leading to misclassification as Case 3 instead of Case 2.

\begin{table}[ht]
\centering
\scriptsize
\begin{adjustbox}{width=\columnwidth}
\begin{tabular}{lrrrrrr}
\toprule
\textbf{Attack} & \textbf{UWave} & \textbf{ERing} & \textbf{BasicMotions} & \textbf{Cricket} & \textbf{NATOPS} & \textbf{AWR} \\
\midrule
FGSM     & \textbf{100.0/ 100.0} & \textbf{100.0}/ 97.8 & \textbf{98.0}/ 97.5 & 0.0/ \textbf{100.0} & \textbf{100.0}/ 61.1 & \textbf{100.0/ 100.0} \\
BIM      & \textbf{100.0/ 100.0} & \textbf{100.0}/ 92.2 & 0.0/ \textbf{100.0} & \textbf{100.0/ 100.0} & \textbf{96.7}/ 50.6 & \textbf{100.0/ 100.0} \\
DeepFool & \textbf{87.7}/ 41.3 & \textbf{100.0}/ 57.0 & 0.0/ \textbf{100.0} & \textbf{54.2}/ 30.6 & \textbf{15.6/ 15.6} & 98.7/\textbf{100.0} \\
MIM      & \textbf{100.0/ 100.0} & \textbf{100.0}/ 92.6 & 0.0/ \textbf{97.5} & \textbf{100.0/ 100.0} & \textbf{100.0}/ 61.1 & 98.7/ \textbf{100.0} \\
AutoPGD     & \textbf{100.0/ 100.0} & \textbf{100.0}/ 91.5 & 0.0/ \textbf{100.0} & \textbf{100.0/ 100.0} & \textbf{90.6}/ 48.3 & \textbf{100.0/ 100.0} \\
ElasticNet & \textbf{100.0/ 100.0} & \textbf{100.0/ 100.0} & 0.0/ \textbf{100.0} & \textbf{100.0/ 100.0} & \textbf{100.0/ 100.0} & \textbf{100.0}/ 96.93 \\
Boundary & \textbf{100.0/ 100.0} & \textbf{100.0/ 100.0} & \textbf{100.0/ 100.0} & \textbf{100.0/ 100.0} & \textbf{100.0/ 100.0} & \textbf{100.0}/ 99.67 \\

\bottomrule
\end{tabular}
\end{adjustbox}
\caption{Comparison of Fourier (F) and Wavelet (W) attack detection rates for Case 2. Format: \textit{F / W}, all values in \%. Bold indicates the higher of the two methods.}
\label{tab:fw_case2_comparison}
\end{table}

For Case 3, five random scenarios were evaluated for each dataset in the PBD. The detection behavior in Case 3 largely mirrored the outcomes of Cases 1 and 2.
As shown in Table~\ref{tab:fw_percent_decision_case3}, detection results are reported in the format Fourier/Wavelet (Final Decision), where the final classification is based on the higher of the two scores. The Final Decision, which represents the attack intensity level—20\%, 40\%, 60\%, or 80\%—is estimated by selecting the closest matching level to this maximum score, allowing for a ±10\% tolerance around each target. In BasicMotions, Fourier failed to detect both clean and attacked segments that were successfully identified by Wavelet, leading to correct final decisions. In contrast, DeepFool attacks on UWave and ERing were missed by Wavelet but detected by Fourier. Overall, only DeepFool attacks on NATOPS and Articulatory Word Recognition were not detected by either method, resulting in the only misclassifications in Case 3. This is due to DeepFool's minimal and fine-grained perturbations, which are designed to closely follow the decision boundary and often leave little detectable signal distortion in either the time or frequency domain. All other segments were correctly
classified.

After classifying the incoming data as partially attacked and estimating its attack intensity level as 20\%, 40\%, 60\%, or 80\%, each of the five segments was independently processed through Module 1 to determine whether it was attacked or clean. The attack detection performance at the segment level was consistent with the results observed in Case 2, showing the reliability of the underlying detection mechanism. These segment level predictions were then aggregated into a pattern (such as attacked/clean/attacked/clean/clean) and passed to Module 2 to classify the attack group associated with the attacked segments of the data.

We examine the performance of Module 1 across different domains with two datasets: ArticulatoryWordRecognition (AWR) and UWaveGestureLibrary (UWave), which originate from distinct domains— Motion and HAR, respectively. Both datasets demonstrated reliable performance in Module 1, achieving high attack detection rates across clean, partially attacked, and fully attacked settings, with minimal false positives or negatives. Notably, AWR remained robust even against subtle perturbations, while UWave exhibited near-perfect detection across all adversarial methods, including DeepFool, which was particularly challenging for other datasets.

\begin{table}[ht]
\footnotesize
\centering
\begin{adjustbox}{width=\columnwidth}
\begin{tabular}{lccccc}
\toprule
\textbf{Dataset} & \textbf{Sequence 1} & \textbf{Sequence 2} & \textbf{Sequence 3} & \textbf{Sequence 4} & \textbf{Sequence 5} \\
\textbf{} & \multicolumn{5}{c}{\textit{Format: F / W (D), all in \%}} \\
\midrule
NATOPS & \textbf{63}/ 44 (60) & \textbf{28/ 28} (20) & \textbf{47}/ 37 (40) & 33/ \textbf{44} (40) & \textbf{64}/ 45 (60) \\
UWave & \textbf{63/ 63} (60) & \textbf{27}/ 23 (20) & \textbf{80}/ 77 (80) & \textbf{44}/ 43 (40) & \textbf{82}/ 71 (80) \\
Cricket & 61/ \textbf{63} (60) & \textbf{24/ 24} (20) & \textbf{60}/ 50 (60) & 42/ \textbf{43} (40) & \textbf{72}/ 68 (80) \\
ERing & \textbf{65}/ 63 (60) & 30/ \textbf{34} (40) & \textbf{82}/ 57 (80) & \textbf{48}/ 46 (40) & \textbf{83}/ 81 (80) \\
BasicMotions & 0/ \textbf{63} (60) & 0/ \textbf{30} (20) & 0/ \textbf{80} (80) & 0/ \textbf{45} (40) & 0/ \textbf{85} (80) \\
AWR & \textbf{62}/ 45 (60) & \textbf{23/ 23} (20) & \textbf{80/ 80} (80) & \textbf{42/ 42} (40) & \textbf{81/ 81} (80) \\
\midrule
Ground Truth & \textbf{60} & \textbf{20} & \textbf{80} & \textbf{40} & \textbf{80} \\
\bottomrule
\end{tabular}
\end{adjustbox}
\caption{Comparison of Fourier (F) and Wavelet (W) detection rates relative to Decision (D) baseline for each sequence. Format: F / W (D), all values in \%. Bold indicates the higher value between Fourier and Wavelet.}
\label{tab:fw_percent_decision_case3}
\end{table}

\subsubsection{Module 2: Adversarial Attack Group Classification Results}
The aim of Module 2 is to classify the data flagged as attacked in Module 1 into one of two attack groups: Group 1: Iteration- Based Attacks or Group 2: Optimization- and Decision- Based Attacks. Since Case 1 corresponds to clean (unattacked) data, no experiments are performed for that scenario. For Case 2, all attack scenarios are evaluated. In Case 3, only the segments identified as attacked by Module 1 are isolated and evaluated. The results of these experiments are summarized in Table~\ref{tab:performance_module2_case2_by_dataset} for Case 2 and Table~\ref{tab:performance_module2_case3_by_dataset} for Case 3. According to the Table~\ref{tab:performance_module2_case2_by_dataset}, the mean classification accuracy is 92.21\% for Group 1 and 84.93\% for Group 2, yielding an overall average accuracy of 88.69\% for Case 2. The classification performance on isolated segments in Case 3 closely mirrors the results observed in Case 2, though with slightly lower overall accuracy as shown in Table~\ref{tab:performance_module2_case3_by_dataset}. This consistency suggests strong alignment between the two cases, as the segments analyzed in Case 3 are essentially smaller portions of the same datasets used in Case 2, and the detected attacked segments retain similar characteristics to the fully attacked data.

\begin{table}[ht]
  \centering
\begin{tabular}{lccc}
  \toprule
  \textbf{Dataset} & \makecell{\textbf{Average}\\\textbf{Accuracy}} & \makecell{\textbf{Group 1}\\\textbf{Accuracy}} & \makecell{\textbf{Group 2}\\\textbf{Accuracy}} \\
  \midrule

    \midrule
    NATOPS                           & 90.26    & 97.41   & 84.90    \\
    UWaveGestureLibrary              & 85.67    & 88.71   & 81.61    \\
    Cricket                          & 87.28    & 90.08   & 83.55    \\
    ERing                            & 90.57    & 95.80   & 83.60    \\
    BasicMotions                     & 88.43    & 89.17   & 87.88    \\
    ArticularyWordRecognition       & 89.90    & 92.11   & 88.06    \\
    Average       & 88.69    & 92.21   & 84.93    \\
    \bottomrule
  \end{tabular}
  \caption{Classification accuracy and attack‐group performance for each dataset for Case 2}
  \label{tab:performance_module2_case2_by_dataset}
\end{table}

\begin{table}[ht]
  \centering
\begin{tabular}{lccc}
  \toprule
  \textbf{Dataset} & \makecell{\textbf{Average}\\\textbf{Accuracy}} & \makecell{\textbf{Group 1}\\\textbf{Accuracy}} & \makecell{\textbf{Group 2}\\\textbf{Accuracy}} \\
  \midrule

    \midrule
    NATOPS                           & 92.06    & 97.51   & 84.80    \\
    UWaveGestureLibrary              & 85.54    & 88.53   & 81.55    \\
    Cricket                          & 87.18    & 89.83   & 83.65    \\
    ERing                            & 90.40    & 95.49   & 83.62    \\
    BasicMotions                     & 88.55    & 89.15   & 87.75    \\
    ArticularyWordRecognition       & 89.22    & 92.02   & 88.15    \\
    Average       & 88.83    & 92.08   & 84.92    \\
    \bottomrule
  \end{tabular}
  \caption{Classification accuracy and attack‐group performance for each dataset for Case 3}
  \label{tab:performance_module2_case3_by_dataset}
\end{table}

\begin{table}[]
\centering
\caption{ReLATE+ results for Case 1 using accuracy}
\label{tab:ReLATE+_Accuracy}
\resizebox{\columnwidth}{!}{ 
\label{table:case_comparison}
\begin{tabular}{cccccc}
\hline
\textbf{Case}   & \textbf{Oracle} & \textbf{ReLATE+} & \textbf{\begin{tabular}[c]{@{}c@{}}Random Model\\ Selection\end{tabular}} & \textbf{\begin{tabular}[c]{@{}c@{}}Worst Model\\ Performance\\ \end{tabular}} \\ \hline
\textbf{Case 1} & 95.00           & 93.45            & 81.00                                                                      & 44.38                                                                   \\ \hline
\end{tabular}
}
\end{table}
To assess Module 2 across domains, we evaluate AWR (speech) and UWave (gesture). Both datasets demonstrated high classification accuracy when distinguishing between Group 1 and Group 2, effectively capturing the underlying attack group structures. Their performance remained consistent even when isolating and classifying individual attacked segments, indicating robust generalization at the segment level. While both performed well, AWR slightly outperformed UWave overall, particularly excelling in Group 2 scenarios, showcasing stability across varied segment compositions and reliability under more complex, mixed-attack conditions.

\subsubsection{Module 3: Most Resilient Model Selection Results} 
\begin{figure}[t]
    \centering
    \includegraphics[width=0.48\textwidth]{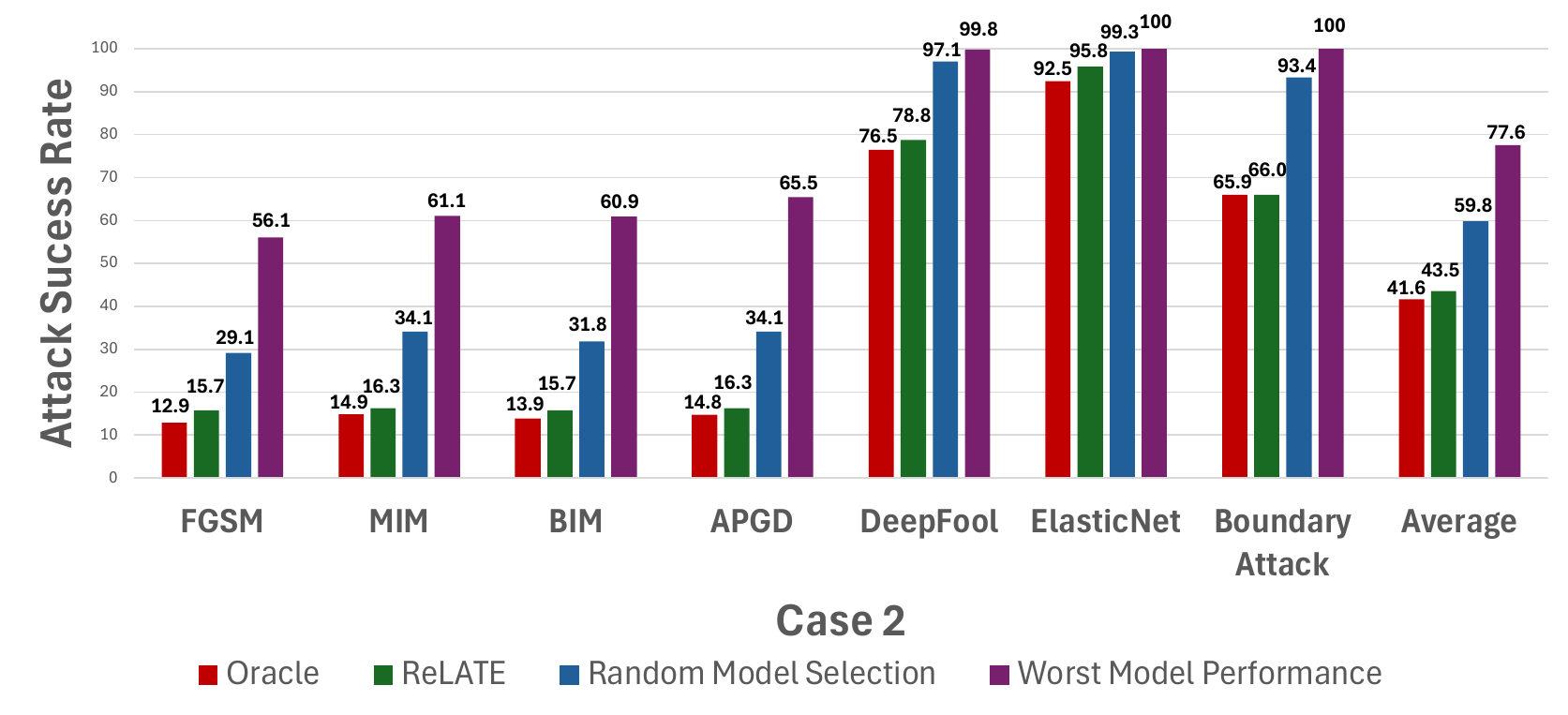}
    \caption{ReLATE+ Case 2 ASR results}
    \label{fig:ReLATE+_ASR_Case_2}
\end{figure}

\begin{figure}[t]
    \centering
    \includegraphics[width=0.48\textwidth]{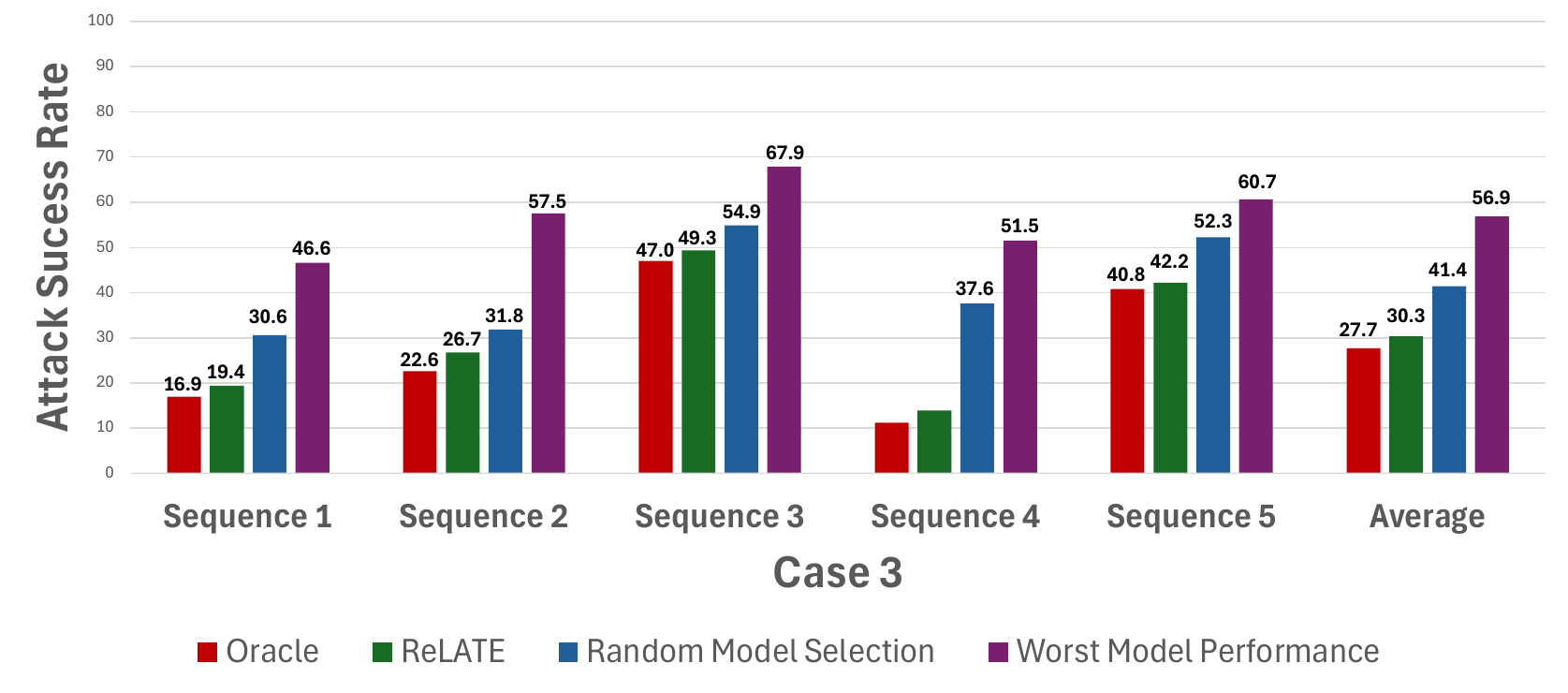}
    \caption{ReLATE+ Case 3 ASR results}
    \label{fig:ReLATE+_ASR_Case_3}
\end{figure}

The goal of this module is to select the most robust model for the incoming data by identifying the most similar dataset in the PBD and transferring the best-performing resilient algorithm associated with that dataset. Table~\ref{tab:ReLATE+_Accuracy}, Figure~\ref{fig:ReLATE+_ASR_Case_2}, and Figure~\ref{fig:ReLATE+_ASR_Case_3} present the results for Module 3. All results are averaged across datasets. The performance of ReLATE+ is compared against Oracle, random model selection, and the worst-performing model. Table~\ref{tab:ReLATE+_Accuracy} shows ReLATE+’s performance in Case 1 (i.e., under no adversarial attack). Oracle refers to the highest accuracy achieved for each dataset, while random selection reflects the average accuracy across multiple randomized model selections. The worst-performing model corresponds to the lowest accuracy per dataset. ReLATE+ achieves accuracy close to Oracle, with only a 1.55\% gap, and outperforms random model selection by 12.45\%.  These results demonstrate ReLATE+’s ability to match datasets with appropriate models based on similarity, without the need for exhaustive testing.


Figure~\ref{fig:ReLATE+_ASR_Case_2} shows results for Case 2 across all attack types, along with their average performance across datasets, providing a comprehensive assessment of ReLATE+’s robustness under different adversarial conditions. Figure~\ref{fig:ReLATE+_ASR_Case_3} presents detailed results for each of the five randomized attack sequences in Case 3. The repetition of Case 3 five times accounts for the variability introduced by randomized attacks and evaluates ReLATE+’s consistency across diverse scenarios. In both cases, Oracle (red) is defined as the model with the lowest ASR for each dataset, while ReLATE+ (green) represents the proposed approach. Random selection (blue) performance is computed using a Monte Carlo approach, while the worst model (purple) corresponds to the highest average ASR. In Case 2, ReLATE+’s ASR is, on average, only 1.9\% higher than Oracle's, yet 16.3\% lower than that of random model selection, and 17.8\% lower than the worst-performing model. Similarly, in Case 3, ReLATE+ exceeds Oracle's ASR by just 2.6\%, while outperforming random selection and the worst model by 11.1\% and 15.5\%, respectively. This underscores ReLATE+'s ability to utilize dataset-specific similarities to recommend models with enhanced adversarial robustness. By aligning model recommendations with the feature distributions of datasets, ReLATE+ effectively mitigates the impact of adversarial attacks.

Overall, ReLATE+ performs consistently close to the Oracle in all cases, even when the performance gap between the worst and best-performing models is substantial. On average across Case 1, Case 2, and Case 3, ReLATE+ is within 2.02\% of the Oracle, achieves a 13.28\% improvement over random model selection, and outperforms the worst-performing model by 27.32\%. This highlights ReLATE+’s ability to achieve near-optimal performance and robustness across diverse adversarial scenarios, all while significantly reducing computational cost by avoiding exhaustive model testing.

Analysis on AWR and UWave consistently showed strong model selection results with ReLATE+, performing close to the oracle and outperforming both random and worst-case baselines. In Case 1, AWR nearly matched the oracle (99.00\% vs. 99.33\%), and UWave also performed competitively (89.44\% vs. 93.89\%). In Case 2, both achieved significantly lower ASRs than the baselines, with AWR at 23.6\% and UWave at 42.9\%, close to their respective oracle-level performances. In Case 3, averaged over five randomized sequences, ReLATE+ maintained low ASRs for both datasets—23.6\% for AWR and 29.7\% for UWave—remaining well below random and worst-case outcomes. These results highlight ReLATE+’s ability to generalize under varying attack patterns and reliably select robust models across domains.



\subsection{Overhead Analysis}
In the absence of ReLATE+,  to achieve optimal model performance (Oracle) all DL models have to be trained. Thus, Oracle overhead is determined by evaluating all models for each case. This process involves training each model and applying all adversarial attacks, ensuring a comprehensive assessment of their performance. Since ReLATE+ focuses on choosing the most similar dataset and evaluating only the top three models, it significantly reduces computational overhead. All overhead values reported are based on execution time measurements. In Module 1, its overhead is negligible compared to the Oracle. In Module 2, ReLATE+ requires only 3\% of the Oracle’s cost to prepare the training data, train an XGBoost model, and perform prediction. In Module 3, it reduces model training and evaluation overhead by 85.12\% in Case 1, 77.26\% in Case 2, and 79.25\% in Case 3, resulting in an average reduction of 80.54\% across all cases. The additional cost of computing the similarity metric is minimal, contributing just 1.02\%, 1.05\%, and 0.53\% of Oracle overhead in Cases 1, 2, and 3, respectively. When comparing total overhead, including all three modules, ReLATE+ achieves a reduction of 81.07\% in Case 1, 73.21\% in Case 2 and 75.73\% in Case 3. Even when including ReLATE+’s internal framework costs, the overall reduction remains high at X\% on average. These results
highlight how ReLATE+ significantly reduces computational
costs by efficiently selecting resilient models, ensuring both
optimal performance and resource efficiency, even under adversarial conditions.

\begin{figure}[t]
    \centering
    \includegraphics[width=0.48\textwidth]{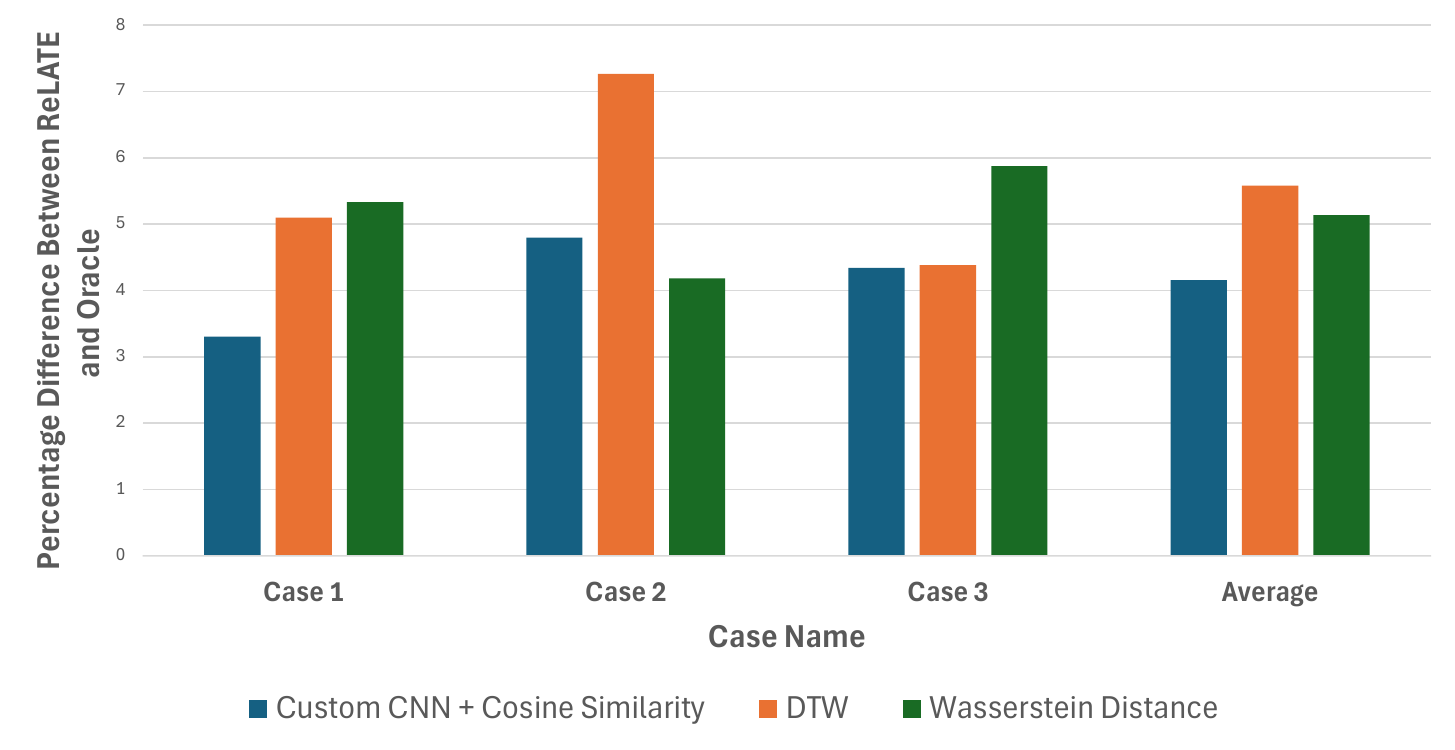}
    \caption{Similarity metric performance comparison}
    \label{fig:Similarity_Metric}
\end{figure}

\subsection{Performance Analysis of Similarity Metrics}
We compare three well-known methods: our approach (CNN + Cosine Similarity), Dynamic Time Warping (DTW)\cite{Petitjean2011}, and Wasserstein Distance \cite{Flamary2017} as dataset similarity metrics, and evaluate their performance across Cases 1, 2 and 3. Figure \ref{fig:Similarity_Metric} shows the average percentage difference between Oracle and ReLATE+ when using each similarity metric. In Case 1, this difference is calculated as the accuracy difference percentage. In all other cases, it represents the ASR percentage difference with respect to the Oracle.
While CNN combined with Cosine Similarity does not achieve the highest performance in every case, it exhibits the smallest average performance deviation and consistently performs near the optimal metric when not the best. It also achieves the highest computational efficiency, reducing computation time by 20\% in Case 1, 15\% in Case 2 and 18\% in Case 3, averaging a 17.7\% improvement over DTW and Wasserstein Distance, which exhibit nearly identical execution times.

\subsection{Ablation Study on Epsilon in Adversarial Attacks}

\begin{table}[ht]
\footnotesize
\centering
\begin{adjustbox}{width=\columnwidth}
\begin{tabular}{lcc}
\toprule
\textbf{Attack} & \textbf{Fourier ($\epsilon = 0.01$ / $0.1$ / $0.2$)} & \textbf{Wavelet ($\epsilon = 0.01$ / $0.1$ / $0.2$)} \\
\midrule
FGSM  & 57.4 / 83.3 / 83.3  & 44.0 / 92.7 / 99.3 \\
BIM   & 57.1 / 82.8 / 83.3  & 44.9 / 90.6 / 100.0 \\
MIM   & 58.4 / 83.9 / 83.3  & 44.7 / 91.9 / 99.6 \\
AutoPGD  & 53.5 / 81.8 / 83.3  & 42.7 / 90.3 / 100.0 \\
\bottomrule
\end{tabular}
\end{adjustbox}
\caption{Adversarial attack detection accuracy (\%) for Fourier and Wavelet methods under different $\epsilon$ values in Case 2, using attack implementations described in~\cite{costa2024deep}.}

\label{tab:epsilon_ablation_results_case2}
\end{table}

To evaluate the sensitivity of ReLATE+ to adversarial perturbation strength, we conducted an ablation study by varying the attack parameter $\epsilon$, which controls the magnitude of perturbation in gradient-based attacks such as FGSM, BIM, AutoPGD, and MIM. While our primary experiments were conducted with $\epsilon = 0.1$, we repeated the analysis using $\epsilon = 0.001$ and $\epsilon = 0.2$ to assess performance under weaker and stronger attack settings respectively. Since Case 1 involves no attacks and Case 3 follows a randomized attack sequence that includes methods not governed by $\epsilon$ (such as DeepFool, ElasticNet, and Boundary Attack), changes in $\epsilon$ cannot be directly isolated in those scenarios. Accordingly, the ablation study was conducted solely on Case 2, which consists of fully attacked scenarios by a variety of adversarial methods; however, only the $\epsilon$-based gradient attacks were considered for this evaluation.

In Module 1, as shown in Table~\ref{tab:epsilon_ablation_results_case2}, the detection performance varied notably with changes in $\epsilon$. These results reflect average detection accuracy across all datasets. When $\epsilon = 0.01$, both Fourier and Wavelet based detection accuracy dropped significantly, indicating that subtle adversarial perturbations are more difficult to detect. In contrast, using $\epsilon = 0.1$ resulted in a sharp increase in detection rates across all attack types. When the attack strength was further increased to $\epsilon = 0.2$, Wavelet based detection improved even further, reaching approximately 100\% accuracy under several attacks. However, Fourier based performance remained largely stable between $\epsilon = 0.1$ and $\epsilon = 0.2$. This stability is primarily attributed to the BasicMotions dataset, where Fourier based detection fails to capture perturbations effectively at both levels of attack strength, thus limiting the average gain in overall performance.

\begin{table}[ht]
\footnotesize
\centering
\begin{adjustbox}{width=\columnwidth}
\begin{tabular}{lccc}
\toprule
\textbf{Dataset} & \textbf{Group 1 Accuracy ($\epsilon = 0.01$)} & \textbf{Group 1 Accuracy ($\epsilon = 0.1$)} & \textbf{Group 1 Accuracy ($\epsilon = 0.2$)} \\
\midrule
UWave                          & 87.68 & 88.71 & 95.80 \\
ERing                          & 91.67 & 95.80 & 95.80 \\
BasicMotions                   & 84.37 & 89.17 & 90.22 \\
Cricket                        & 89.39 & 90.08 & 93.10 \\
NATOPS                         & 88.20 & 97.41 & 98.10 \\
AWR      & 82.27 & 92.11 & 93.05 \\
\midrule
\textbf{Average}               & 87.26 & 92.21 & 94.34 \\
\bottomrule
\end{tabular}
\end{adjustbox}
\caption{Group 1 classification accuracy (\%) across datasets under different perturbation strengths $\epsilon$.}
\label{tab:group1_accuracy_epsilons}
\end{table}

In Module 2, we analyze the impact of perturbation strength on Group 1 classification accuracy across different datasets. Table~\ref{tab:group1_accuracy_epsilons} shows that model performance improves consistently as the attack strength increases from $\epsilon = 0.01$ to $\epsilon = 0.2$. These improvements were observed across all datasets. These findings confirm that ReLATE+ not only detects adversarial perturbations effectively but also maintains strong performance in classifying attack types as the perturbation magnitude increases. Lastly, in Module 3, we examine the effect of $\epsilon$ on resilient model selection using dataset similarity. ReLATE+ performance decreased by 0.42\% for $\epsilon = 0.001$, increased by 0.35\% for $\epsilon = 0.2$ compared to the baseline $\epsilon = 0.1$.



\section{Conclusion}

The dynamic nature of time-series data makes it challenging to determine whether inputs are clean, adversarial, or incomplete. Deep learning models are well-suited to address this complexity, as they can learn temporal patterns directly from raw sequences. However, frequent retraining can be impractical due to high computational costs, especially in real-time settings with limited data. These challenges are further compounded by vulnerability to adversarial attacks, where even small perturbations can cause significant misclassifications.
To address these issues, we propose ReLATE+, a resilient learner selection mechanism that integrates attack
detection, attack group classification, and adaptive defense for time-series classification. It uses a three-stage detect–classify–select pipeline to avoid costly model retraining by dataset-level similarity. 
ReLATE+ flags adversarial inputs using Fourier and Wavelet features, classifies attacks with XGBoost, and selects top-performing models from similar datasets for resilient model
selection. Experiments show it reduces computation by 77.68\%, maintains accuracy within 2.02\% of an oracle, and outperforms random selection by 13.28\% on average.



\section*{Acknowledgments}
This work has been funded in part by NSF, with award numbers \#1826967, \#1911095, \#2003279, \#2052809, \#2100237, \#2112167, \#2112665, and in part by PRISM and CoCoSys, centers in JUMP 2.0, an SRC program sponsored by DARPA.

\bibliographystyle{IEEEtran}
\bibliography{biblio}

\end{document}